\documentclass[aip,jmp,preprint,amsmath,amssymb]{revtex4-1}
\usepackage{graphicx}% Include figure files
\usepackage{dcolumn}% Align table columns on decimal point
\usepackage{bm}% bold math
\usepackage[utf8]{inputenc}
\usepackage[T1]{fontenc}
\usepackage{mathptmx}
\usepackage{etoolbox}
\usepackage{xcolor}

\makeatletter
\def\@email#1#2{%
 \endgroup
 \patchcmd{\titleblock@produce}
  {\frontmatter@RRAPformat}
  {\frontmatter@RRAPformat{\produce@RRAP{*#1\href{mailto:#2}{#2}}}\frontmatter@RRAPformat}
  {}{}
}%
\makeatother
\begin{document}

\preprint{AIP/123-QED}
\title{Vibrational resonance in the FitzHugh-Nagumo neuron model under state-dependent time delay }

\author{M. Siewe Siewe}
\affiliation{University of Yaounde I, Faculty of Science, Department of Physics, Laboratory of Mechanics, Materials and Structures; PO. Box: 812 Yaounde-Cameroon }

\author{S. Rajasekar}
\affiliation{School of Physics, Bharathidasan University, Tiruchirapalli 620024, Tamilnadu, India}

\author{Mattia Coccolo$^*$}
\affiliation{Nonlinear Dynamics, Chaos and Complex Systems Group, Departamento de F\'{i}sica, Universidad Rey Juan Carlos, Tulip\'{a}n s/n, 28933 M\'{o}stoles, Madrid, Spain}\email[Corresponding author: ]{mattiatommaso.coccolo@urjc.es.}

\author{Miguel A.F. Sanju\'{a}n}
\affiliation{Nonlinear Dynamics, Chaos and Complex Systems Group, Departamento de F\'{i}sica, Universidad Rey Juan Carlos, Tulip\'{a}n s/n, 28933 M\'{o}stoles, Madrid, Spain}

\date{\today}% It is always \today, today,

\begin{abstract}
We propose a nonlinear FitzHugh--Nagumo neuronal model with an asymmetric potential driven by both a high-frequency and a low-frequency signal.  Our numerical analysis focuses on the influence of a state-dependent time delay on vibrational resonance and delay-induced resonance phenomena. The response amplitude at the low-frequency is explored to characterize the vibrational resonance and delay-induced resonance. Our results show that for smaller values of the amplitude of the state-dependent time-delay velocity component, vibrational resonance and multi-resonance occur in the neuronal model. For large values of the high-frequency excitation amplitude, vibrational resonance appears with one peak. Furthermore, we observe a change in the response when the amplitude of the state-dependent time-delay velocity component increases. In addition, we analyze how the state-dependent time-delay position and velocity components can give birth to delay-induced resonance for separate and together. The key findings of this work demonstrate that the state-dependent time-delay velocity component plays a crucial role in both phenomena. Specifically, the delay parameter serves as a critical control factor, capable of triggering the onset of the two resonances.
\end{abstract}

\maketitle

\begin{quotation}
This work examines a FitzHugh-Nagumo neuronal model with an asymmetric potential subjected to dual-frequency excitation, combining high- and low-frequency signals. Emphasis is placed on the role of state-dependent time delays in shaping vibrational and delay-induced resonance phenomena. Through numerical simulations, the resonance effects are evaluated by analyzing the response amplitude to the low-frequency signal. The findings indicate that smaller velocity components of the state-dependent time delay give rise to both single and multiple vibrational resonances, while larger amplitudes of high-frequency excitation primarily produce a single-peak vibrational resonance. Moreover, increasing the velocity component of the time delay diminishes the system's response. The investigation also uncovers how the position and velocity components of the state-dependent delay, acting independently or in tandem, contribute to delay-induced resonance, offering a deeper understanding of the dynamic interactions in neuronal systems with state-dependent feedback.
\end{quotation}

\section{Introduction}

The modeling of neuronal dynamics is a significant area of research within Nonlinear Dynamics and Chaos, with applications to biological systems. One of the pioneering contributions to this field was carried out by Hodgkin and Huxley, who described the relationship between action potentials and the ion currents \cite{b1,b2}. Due to the complexity of analyzing this system, simpler models were developed to provide a more accessible approach. Among the most notable low-dimensional models are the FitzHugh-Nagumo \cite{b3}, Morris-Lecar \cite{b4} and Hindmarsch-Rose \cite{b5,b6} models, among others. These models are continuous-time systems, modeled by differential equations, and are well established in the literature \cite{b7}.

Over the last three decades, vibrational resonance, a type of nonlinear resonance first reported by Landa and McClintock \cite{a4}, has received great attention \cite{c0,c1,Rev_Sanjuan}. Vibrational resonance occurs in nonlinear systems driven by a low-frequency force, $f \cos \omega t$, and a high-frequency force, $g \cos \Omega t$, where $\Omega \gg \omega$. Essentially, an optimal high-frequency driving amplitude enhances the system's response to a low-frequency subthreshold signal. The difference with respect to the models presented in \cite{Rev_Sanjuan} is that our FitzHugh-Nagumo model employs state-dependent delays that adapt dynamically based on the system's state (e.g., membrane potential $x$ or its derivative $\dot{x}$), enabling more biologically realistic feedback mechanisms. In contrast, many models reviewed in \cite{Rev_Sanjuan} utilize fixed or externally imposed delays, which are less adaptable and do not reflect the intrinsic state of the system. Our model emphasizes its biological relevance, particularly in simulating neuronal dynamics, where state-dependent memory and feedback are critical. In contrast, models like the Van der Pol oscillator are primarily focused on general oscillatory behavior in physical or engineering systems, often lacking the biological context our FitzHugh-Nagumo model addresses. This phenomenon has been extensively studied numerically, analytically, and experimentally \cite{c0,c1,a5,a6,a7,a8,a9}.

Often, it is assumed that the time-dependent system under consideration satisfies the Markov property, meaning that the future states of the system are entirely determined by the current state and are independent of the past. In such cases, the system can be adequately described by ordinary or partial differential equations. However, the principle of Markovianity is frequently only a first approximation of the true situation, and a more realistic model would incorporate past states of the system \cite{a10}. Systems with state-dependent delays arise when the dynamics depend not only on the instantaneous configuration but also on a delayed configuration, with the delay size being contingent on the state of the system. In many applications, these delays are time-varying or state-dependent. A significant area of study for state-dependent time delays is in neuronal dynamics, where signal propagation often involves delays that vary with the system's state.

Time-delay effects have been incorporated into the study of vibrational resonance. The delay parameter has been observed to induce periodic vibrational resonance in simple delayed bistable systems, where the period of the resonance peak relative to the time-delay parameter matches the period of the high-frequency excitation signal, as demonstrated in prior studies \cite{vasilenko2012,liu2014,giacomelli2006}. Vibrational resonance with time delay has also been analyzed in the FitzHugh-Nagumo system \cite{a12}. Most previous research on vibrational resonance focused on the effects of time delay with a single constant, multiple constants, integrative over a finite interval, and distributive over an interval with specific distributions.
The primary distinction between the vibrational resonance observed in the studied FitzHugh-Nagumo model with state-dependent time delays and the traditional FitzHugh-Nagumo model with fixed delays lies in the adaptability of the delay mechanism. In the classical model, fixed delays represent constant feedback or communication lags, capturing basic feedback effects but lacking the adaptability characteristic of biological systems. In contrast, the state-dependent delays in the studied FitzHugh-Nagumo model vary dynamically with the system's state, such as the membrane potential or its derivative. This dynamic coupling introduces a more biologically plausible feedback mechanism, allowing for the modeling of complex neural processes, including adaptive feedback, memory effects, and synchronization phenomena. Such state-dependent delays lead to richer dynamical behaviors, such as delay-induced oscillations, which are crucial for accurately simulating the behavior of neural circuits.

Incorporating state-dependent delays in both the position (\(x\)) and velocity (\(\dot{x}\)) variables is essential for capturing a broad spectrum of biologically relevant feedback mechanisms. This dual inclusion reflects the complexity of real-world neural systems, where feedback mechanisms often affect multiple state variables simultaneously. By introducing delays into both components, the model can more faithfully reproduce phenomena such as vibrational resonance. The interaction between these delayed components is crucial for accurately modeling the dynamic coupling that underlies the observed resonance effects. Limiting the delay solely to the velocity term would oversimplify the system, potentially omitting critical interactions between the position and velocity, and highlighting the importance of this comprehensive approach.

However, vibrational resonance of the FitzHugh-Nagumo neuron model with bistability under the effect of state-dependent time delay has also attracted much attention due to its simplicity and
flexibility. In the Ref.~\cite{a14}, a time delay described by a sigmoid function, a parabolic function, and a Gaussian function of position variable are considered. Nontrivial effects are realized in the presence of 
position-dependent time-delayed feedback. Parabolic state-dependent time delay is found to give rise to the occurrence of multiple vibrational resonances for a range of
some parameters. Motivated by the concept of impulsive resets of state variables, specifically state-dependent resets caused by an action potential \cite{b19}, and the reported effects of state-dependent time delays \cite{a14}, we explore a modified FitzHugh-Nagumo neuron model. This model incorporates nonlinear damping, an asymmetrical potential \cite{a15,a16}, and is subjected to state-dependent time delays and a biharmonic force.

A notable phenomenon that arises from the delay is the {\it delay-induced resonance}, where the system exhibits resonance-like behavior due to the intrinsic oscillations caused by the time delay. Delay-induced resonance can be viewed as an extension of vibrational resonance, where oscillations induced by state-dependent delays act as a high-frequency forcing, interacting with external low-frequency forcing to produce resonance phenomena similar to classical vibrational resonance. This view, supported by Lv et al.~\cite{lv2015high}, suggests that the time delay can substitute the high-frequency forcing in driving resonance. In this study, we examine how state-dependent delays in both position and velocity components of the FitzHugh-Nagumo model influence resonance behavior, shedding light on how neural systems process information and possibly contributing to pathological conditions such as epilepsy.

The study of state-dependent time delays and delay-induced resonance \cite{Cantisan,Coccolo} in the FitzHugh-Nagumo system not only advances our understanding of neuronal behavior but also has broader implications for designing control strategies in engineering systems and understanding delay effects in other biological contexts. By delving into the intricate interplay between delays and system states, researchers can uncover fundamental principles that govern the dynamics of delayed feedback systems.

The insights gained from studying state-dependent time delay and delay-induced resonance in the FitzHugh-Nagumo system have several practical applications across various fields. In neuroscience, understanding how delays influence neuronal dynamics can help in developing more effective treatments for neurological disorders such as epilepsy, where abnormal timing of neuronal firing is a key characteristic. Additionally, these findings can improve the design of artificial neural networks and neuroprosthetic devices by incorporating more accurate models of neuronal signal transmission. Beyond biology, the principles derived from this research can be applied to optimize feedback control systems in engineering, enhance signal processing algorithms, and improve the stability and performance of complex networks such as power grids and communication systems. By leveraging the knowledge of how state-dependent delays affect system behavior, engineers and scientists can develop more robust and efficient systems across a wide range of disciplines.

In summary, we have incorporated both a state-dependent time-delay position component and a state-dependent time-delay velocity component into the FitzHugh-Nagumo neuronal model. Our analysis focused on the effect of the velocity component on two resonance phenomena: vibrational resonance and delay-induced resonance. Ultimately, we found that the parameters of the time-delay velocity component can control the appearance of vibrational resonance, and delay-induced resonance.

The organization of this paper is as follows. In Sec.~\ref{Sec.2}, we give a brief description of the system with model equation and applications. In Sec.~\ref{Sec.3}, the state-dependent time delay
on the vibrational resonance system performance is investigated numerically. The study of the delay-induced resonance in the FitzHugh-Nagumo system end the effect of the state-dependent delay velocity component on the phenomenon is presented in Sec.~\ref{Sec.4}. The main conclusions and a discussion of our results are provided in Sec.~\ref{Sec.5}.

\section{Model description }\label{Sec.2}
\noindent

A considerable number of FitzHugh-Nagumo neuron models
have been developed in the literature \cite{a01,a02}. The study of vibrational resonance in the presence of state-dependent time delay has attracted significant attention.  The main challenge lies in finding a model that accurately describes the range of potential differences across the cell membrane, where vibrational resonance under state-dependent time-delay position and velocities occurs.

\begin{figure}[htp]
\begin{center}
\includegraphics[width=10cm, ,clip=true]{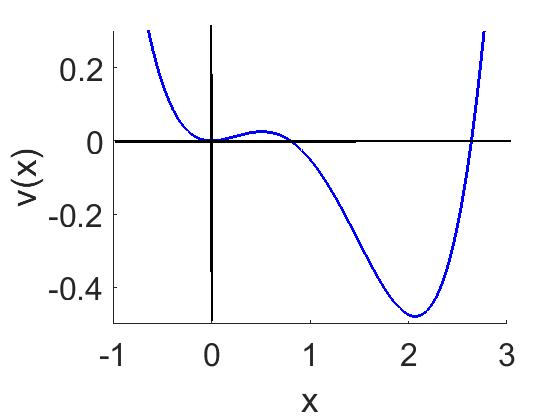}
\caption{The figure illustrates a sketch of the asymmetric potential considered in our FitzHugh-Nagumo model. } \label{Fig.1}
\end{center}
\end{figure}

We used the work by Fitzhugh \cite{a03} and also the references \cite{a04,a05}. Since the  FitzHugh-Nagumo is modelled by coupled equations, we consider that the delay may affect the position and the velocities of the fast variable inside the equation describing the slow variable and given by the following system:
\begin{align}
\label{siewe8}
\dot{x} & =  F(x) - \eta, \\
\dot{\eta} & = x-a_{1}\eta+\gamma x\left(t-\tau(x(t))\right) + \alpha\dot{x}\left(t-\tau(x(t))\right)+f\cos{\left(\omega
t\right)}+g \cos{\left(\Omega t\right)},\nonumber
 \end{align}
where the dot indicates differentiation with respect to $t$, $x(t)$ is the fast variable representing the voltage across a neural membrane in the neural case and $\eta(t)$ is the slow recovery variable related to the time dependent conductance of the potassium channels in the membrane. Likewise, the function $F(x)$ represents the ionic current and may be considered to be in the quadratic form \cite{a04}, $\gamma$ and $\alpha$ are the strength of the position and velocities state-dependent time-delay feedback, $a_{1}$ is a constant and $\tau(x(t))$ is the state-dependent time-delay variable. Here, we assume $\Omega > > \omega$, which means that $gcos (\Omega t)$ and $f cos (\omega t)$ are respectively the high-frequency signal with the amplitude $g$ and the low-frequency signal with the amplitude $f$. The recovery or hyperpolarization of the neuronal action potential may cause communication delays in the neurons. In our  FitzHugh-Nagumo model \cite{a15,a16}, this recovery of the neuronal action potential is a function of the position and velocity of the voltage across a neural membrane, inducing state-dependent time delay in both linear position and velocity terms. The externally applied current is given by $f\cos \omega t + g \cos \Omega t$. Thus,  after eliminating the slow variable in the equations system \ref{siewe8}, one obtain the modified FitzHugh-Nagumo neuron model below.

\begin{equation}
\begin{array}{lcl}
\ddot{x}+\mu
\left(1-0.54x+0.57 x^{2}\right)
\dot{x}+\displaystyle\frac{dV(x)}{dx}+ \gamma x\left(t-\tau(x(t))\right)+\alpha\dot{x}\left(t-\tau(x(t))\right)\\
\\
\quad \quad =f\cos\omega t+g\cos\Omega t,
\end{array}
\label{eq8}
\end{equation}
where $x$ refer to the membrane potentials of the neurons,
$\tau\left(x(t)\right)$ is state-dependent time delay,
$\mu,\gamma,\alpha,f$ and $g$ are parameters. Throughout the text, we fix the parameters $\gamma= 0.2, f=0.1, \omega=1, \Omega=10$. The asymmetrical double-well potential $V(x)$ is defined by:

\begin{equation}
\begin{array}{lcl}
V(x)= \displaystyle\frac{\omega_{0}^{2}}{2}x^{2}-\displaystyle\frac{\lambda}{3}x^{3}+\displaystyle\frac{\beta}{4}x^{4},
\end{array}
\label{eq9}
\end{equation}
with fixed parameters $\omega_{0}=0.81,$ $\lambda=1.60$ and $\beta=0.62$,  as shown in Fig.~\ref{Fig.1}.

This modified FitzHugh-Nagumo model incorporates state-dependent delays in both the position and velocity terms. These delays depend on the system's state, such as the membrane potential \( x \) or its derivative \( \dot{x} \), offering a more accurate and biologically plausible representation. Neural feedback mechanisms often rely on memory that is state-dependent, making the inclusion of such delays essential for modeling complex behaviors. This approach captures feedback effects inherent in biological systems, where synchronization and memory are critical to processes such as pattern formation and neural communication \cite{Tass2005, Matsumoto2006}. 

Notably, the modified FitzHugh-Nagumo model shares mathematical and conceptual parallels with the Van der Pol oscillator, a classical nonlinear dynamical system known for its self-sustained oscillations and resonance phenomena. Like the Van der Pol oscillator’s nonlinear damping term \((\mu - \beta x^2)\dot{x}\), which alternates between dissipative and amplifying regimes, the modified FitzHugh-Nagumo model includes a state-dependent damping term \(\mu(1 - 0.54x + 0.57x^2)\dot{x}\) that governs the system's oscillatory behavior. Additionally, while the classical Van der Pol oscillator lacks inherent delays, time-delayed feedback has been explored in its extended versions. The state-dependent delays \(\gamma x(t-\tau(x(t)))\) and \(\alpha \dot{x}(t-\tau(x(t)))\) in the FitzHugh-Nagumo model introduce memory effects, creating a dynamic framework for studying synchronization and resonance phenomena. Unlike fixed delays, state-dependent delays introduce a feedback loop that varies with the system’s current state, allowing for more realistic modeling of memory and synchronization phenomena observed in neural systems. Here, we specifically refer to synchronization between the neuron and external driving signals. This enhanced modeling framework provides a better mirror of real-world neural dynamics and offers deeper insight into the role of delayed feedback in neuronal behavior. By enriching the FitzHugh-Nagumo model, these modifications make it a more accurate representation of biological systems and provide a powerful tool for exploring delay-induced and vibrational resonance phenomena, broadening the scope of the model's applicability \cite{Hale1991, Liu2014Ch}.

Moreover, state-dependent time-delay velocity components, especially in the context of delay-induced and vibrational resonance, have practical applications across various fields. In neuroscience and biological systems, understanding these dynamics in neuronal networks can contribute to advances in treating neurological disorders and designing artificial neural networks \cite{Masoller}. In the study of cardiac dynamics, such models can improve pacemaker technology \cite{Glass}. In engineering and control systems, they enhance robotic stability and responsiveness \cite{Kelly}, as well as the safety and performance of autonomous vehicles \cite{Behere}. Furthermore, managing delay-induced resonance can optimize data transmission and distributed system performance in communication networks \cite{Stallings}. In mechanical and structural systems, these components improve vibration control and extend the lifespan of industrial machinery \cite{Preumont}. Additionally, in energy systems, they can optimize wind turbine efficiency \cite{Burton} and stabilize power grids \cite{Kundur}. In summary, integrating state-dependent time-delay velocity components with resonance management enhances the performance, stability, and reliability of a wide range of dynamical systems.

Finally, all simulations were performed using Matlab's {\it ddesd} function with adaptive step size. The results obtained with this integrator have been validated by comparison with those from the Euler method for delay differential equations, showing a good match and demonstrating the robustness of the results irrespective of the numerical tool utilized. As stated in the previous part, the initial transients have been excluded from the analysis, as they have negligible impact on the results after the system reaches the steady-state, which is the primary focus of this study.

Now, we turn our attention to the effects of a state-dependent time delay on the vibrational resonance phenomenon. In our work, we define the state-dependent time delay using a sigmoid function as expressed in Eq.~(1) as:

\begin{equation}
\begin{array}{ll}
\tau\left(x(t)\right)  =
\displaystyle\frac{\tau_{0}}{1+\exp(px)},
\end{array}
\label{f10}
\end{equation}
where $\tau_0>0$ and $p$ denotes two parameters.  The choice of Eq.~(\ref{f10}) is motivated by the work done by Resat Ozgur Doruk et al.~\cite{a38}. Moreover, the use of a sigmoid function in the state-dependent time-delay term of Eq.~(\ref{eq8}) is motivated by its ability to realistically model feedback mechanisms observed in biological systems, particularly within the FitzHugh-Nagumo  neuron model. The sigmoid function is a smooth, saturating nonlinearity that mirrors the threshold-like behavior seen in processes such as synaptic transmission and signal propagation in neurons, making it a biologically plausible choice. The sigmoid function facilitates a dynamic mapping between system states (such as position or velocity) and the time delay, allowing for a more accurate representation of the complex interactions between system dynamics, delayed feedback, and external forcing.

This choice improves the biological realism of the model while avoiding the discontinuities or artifacts that could destabilize the system, which are often introduced by simpler delay functions. The introduction of a state-dependent delay, especially when governed by a sigmoid function, enhances the model's ability to capture intricate dynamical behaviors, such as vibrational resonance. This is particularly important as the interactions between delay, feedback, and external forcing are subtle and require a smooth, continuous transition to maintain stability. In comparison to the traditional FitzHugh-Nagumo model, which does not include delay feedback, the inclusion of state-dependent delays introduces memory effects and nonlinearity, enriching the dynamic framework and allowing for a more realistic simulation of neuronal behavior. This ultimately enables the study of complex resonance phenomena, such as delay-induced and vibrational resonance, which would not be possible in a simpler, memoryless system.

\begin{figure}[htp]
\begin{center}
\includegraphics[width=10cm, height= 8cm]{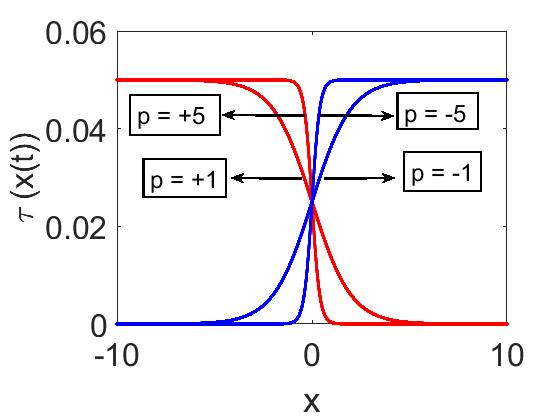}
\caption{The figure shows the graph of the sigmoid function for different values of the parameter $p$. In red for positive values, and in blue for negatives values.} \label{Fig.2}
\end{center}
\end{figure}

The graph of the sigmoid function, shown by the red and blue lines in Fig.~\ref{Fig.2} for different values of $p$, reveals its $S$-shaped curve. As $p$ increases in absolute value, this curve transitions to a $Z$-shaped one. Consequently, an appropriate value for $p$ must be chosen to ensure the model behaves as intended. In all simulations, we take as initial conditions $x_0 = 1.80$ and $\dot{x}_0 = -0.04$, and we expect that our conclusions are of general validity and not specific for the boundary conditions considered. In fact, an additional study examining the impact of varying the initial conditions on the phenomenon was conducted, with no qualitative differences in the outcomes observed.

\section{ Effect of the state-dependent time delays on the vibrational resonance}\label{Sec.3}
\noindent
Usually, the vibrational resonance is characterized by the response amplitude of the system at the low-frequency $\omega$, which is defined by

\begin{equation}
\begin{array}{ll}
Q  =  \sqrt{Q^{2}_{s} + Q^{2}_{c}},
\end{array}
\label{f11}
\end{equation}
where

\begin{eqnarray}
Q_{s} & =  & \frac{2}{mT} \int^{mT}_{0} x(t)\sin\omega t {\mathrm{d}}t \\
\label{f12}
Q_{c} & = & \frac{2}{mT}\int^{mT}_{0}x(t)\cos\omega t {\mathrm{d}} t,
\label{f13}
\end{eqnarray}
and $T = 2\pi/\omega$ is the period of a low-frequency
signal and $m$ is a positive integer. Typically, we
are interested in the transport of the information encoded in the
low-frequency $\omega$, so our focus is in the properties of the parameter $Q$.

\begin{figure}[htp]
\begin{center}
\includegraphics[width=9.9cm, ,clip=true]{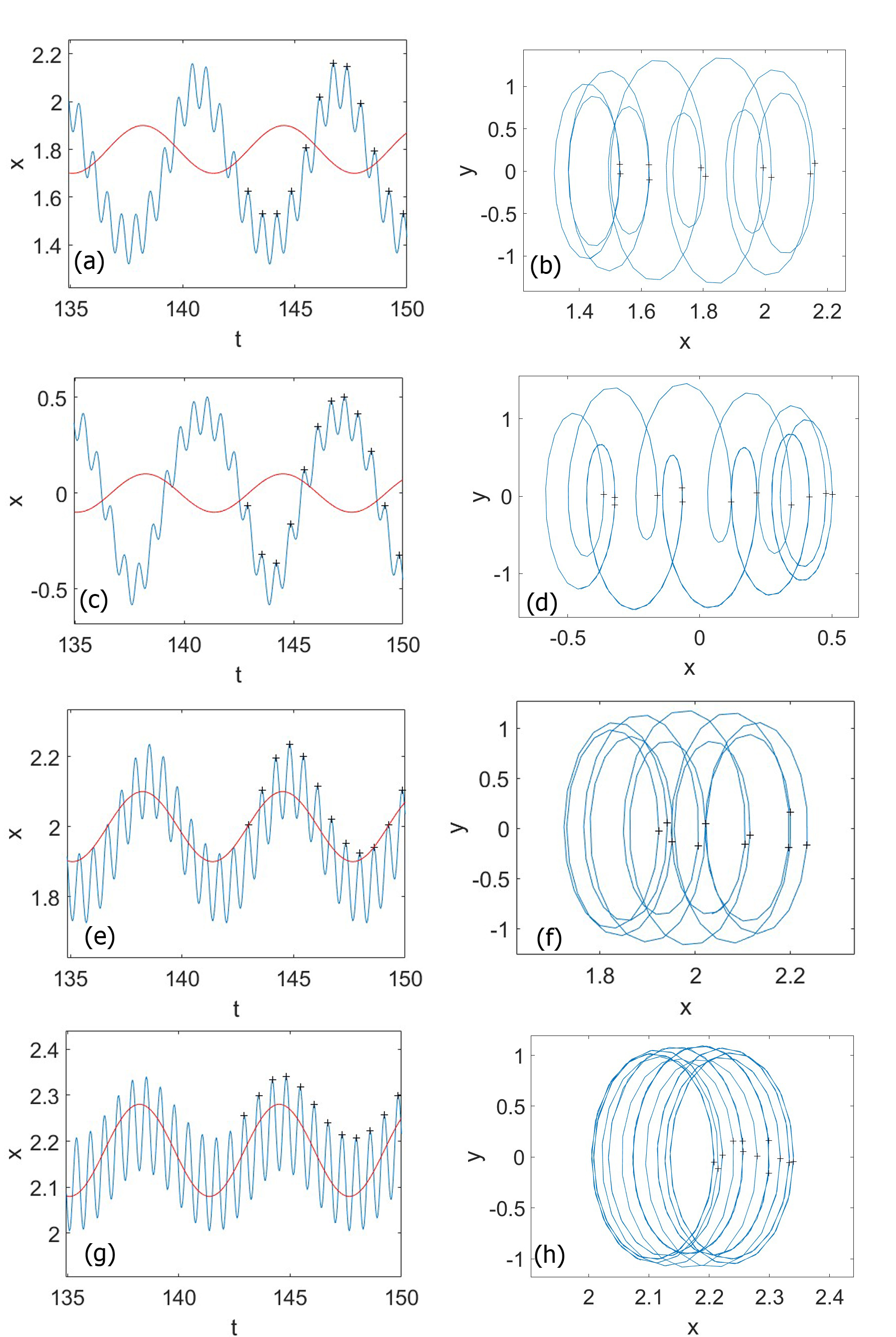}
\caption{Time evolution of the membrane potential \( x(t) \) (blue curve) and the external forcing signal (red curve) for varying values of the state-dependent time-delay velocity component \(\alpha\). The red curve is vertically shifted for clarity. Panels on the left show the temporal dynamics, while panels on the right depict the corresponding phase portraits. Parameters: \(\mu=0.1\), \(g=10\), \(\tau_{0}=0.05\), \(p=-1\). In panels (a) and (c), corresponding to \(\alpha=0.1\) and \(\alpha=5\), respectively, the system exhibits desynchronized oscillations due to weak feedback coupling, as indicated by the phase lag between the blue and red curves. However, in panel (c), interwell oscillations are observed. In panel (e) (\(\alpha=5.4\)), the synchronization is observed, as the blue aligns more closely with the red curve, indicating a better coherence. In panel (f)  (\(\alpha=15\)), full synchronization is achieved with near-perfect overlap of the blue and red curves, demonstrating how increased delay coupling promotes resonance-like behavior and synchronization.} \label{Fig.3}
\end{center}
\end{figure}

\begin{figure}[htp]
\begin{center}
\includegraphics[width=16cm, clip=true]{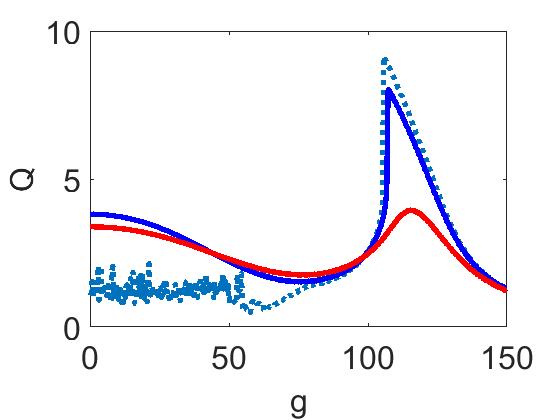}
\caption{ In panel (a) we represent the maxima-minima diagram in function of the parameter $g$ for the case $\alpha=-0.1$. Here, we can see how the dynamics of the system evolve while the $g$ parameter changes. The response amplitude $Q$-factor is plotted in panel (b) as a function of the amplitude of high-frequency force $g$ in the presence of the sigmoidal function with parameters $\mu=0.1$, $f=0.1,$ $\tau_{0}=0.05,$ $p=-1$. The dot line refers to $\alpha=-0.1$, the blue line to $\alpha=-0.01$, and the red line to $\alpha=0.13$.   } \label{Fig.5}
\end{center}
\end{figure}

\begin{figure}[htp]
\begin{center}
\includegraphics[width=13.0cm, ,clip=true]{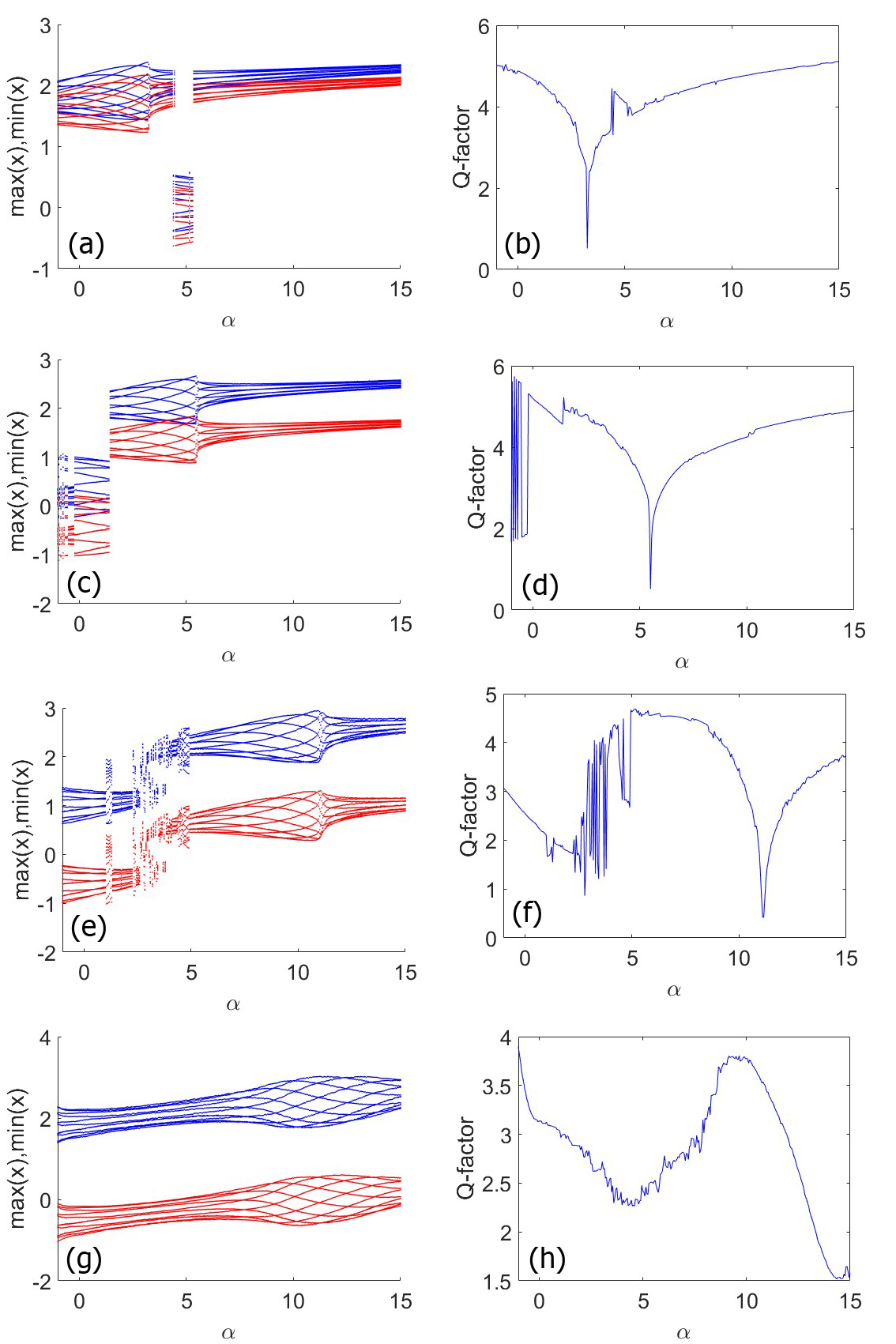}
\caption{ The $Q$-factor and the maxima-minima diagram diagram in function of $-1<\alpha<15$ and for different values of $g$ when $\mu=0.1$, $\tau_{0}=0.05,$ $p=-1$. In panels (a-b) we set $g=10$, in  panels (c-d) $g=40$, in panels (e-f) $g=80$, and in panels (g-h) $g=120$. The $g$ values have been chosen in accord with Fig.~\ref{Fig.5} } \label{Fig.6}
\end{center}
\end{figure}

\begin{figure}[htp]
\begin{center}
\includegraphics[width=16cm, ,clip=true]{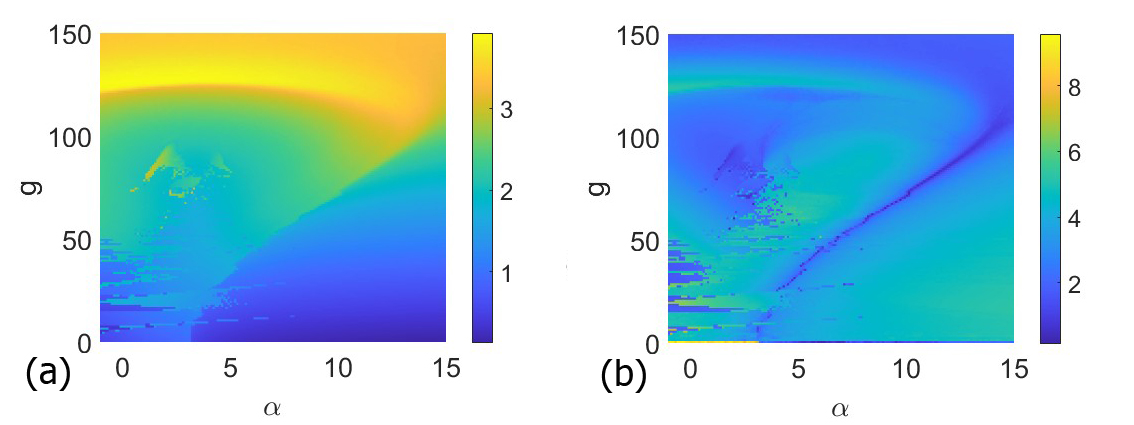}
\caption{ The gradient plots show the set of values of the parameters $g$ and $\alpha$ for which vibrational resonance under the sigmoidal function occurs. Panel (a) shows the oscillation amplitudes and panel (b) the $Q$-factor in the parameter set $\alpha-g$ with $p =1$. } \label{Fig.6b}
\end{center}
\end{figure}

\begin{figure}[htp]
\begin{center}
\includegraphics[width=13.5cm, ,clip=true]{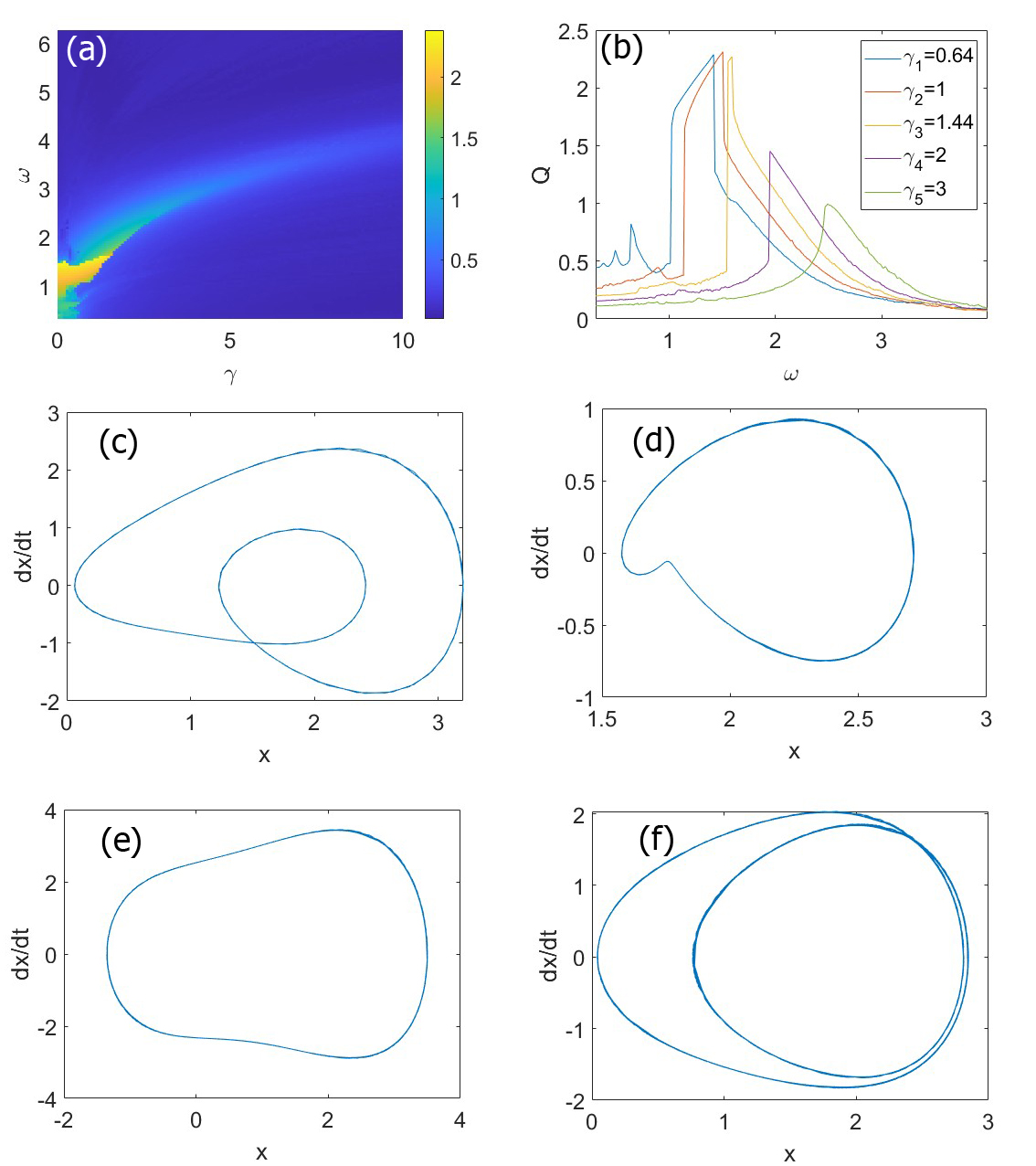}
\caption{ In panel (a) we show the gradient of the $Q-$factor that identifies the presence of the delay-induced resonance in the FitzHugh-Nagumo model with state dependent sigmoidal function time delay in the parameter set of $\gamma$ the multiplier of the position delay and $\omega$ the frequency of the forcing $f\cos{(\omega t)}$. The other parameters are: $\mu=0.1$, $\alpha=0$, $g=0$, $f=1$, $\tau_{0}=0.05,$  $p=-1$. In panel (b) we show three slices of panel (a) for given values of the $\gamma$ parameter, specified in the legend. panels (c)- (f) show the orbits related with the case $\gamma=0.64$, ~$\omega=0.646259,  \omega=1, ~\omega=1.4$ and $\omega=1.5$, respectively. The first frequency value has been chosen to match the small peak, which gives $Q\approx0.8$, the others before, on, and beyond the higher peak.} \label{Fig.7}
\end{center}
\end{figure}

To demonstrate the occurrence of vibrational resonance in a FitzHugh-Nagumo neuron model induced by the state-dependent time-delay velocity component, we fix the amplitudes of the low and high-frequencies signal component and vary the amplitude $\alpha$ of the state-dependent time-delay velocity component. 

The time evolution of the membrane potential $x(t)$ for some specific values of $\alpha$ is depicted in Fig.~\ref{Fig.3}. As anticipated, a more synchronized firing pattern emerges in the neuron as the values of $\alpha$ increase. As a reminder, we refer to synchronization between the neuron and external driving signals. Both strong enough amplitude $\alpha$ of the state-dependent time-delay velocity component and a large enough cluster size are required to achieve regular neuron's firing pattern. This is illustrated in Figs.~\ref{Fig.3}(a), (c), (e) and (g).  We observe in Figs.~\ref{Fig.3}(a) and (c) that synchronization between the fast oscillations and the slower modulation is not achieved, as indicated by the phase lag between the external forcing signal (red curve) and the membrane potential evolution (blue curve). This desynchronization results from the combination of weak coupling (\(\alpha = 0.1\)) and the state-dependent delay term, which introduces phase shifts that hinder coherent alignment of the oscillatory components. Therefore, synchronization in this context refers to the alignment between the neuronal firing pattern and the external driving signals. Notably, panel (c) also exhibits interwell oscillations, highlighting the complex dynamical behavior. 

As the delay parameter increases to \(\alpha = 5.4\) in panel (e), synchronization improves significantly, demonstrating how stronger feedback coupling stabilizes the phase relationship. Numerical experiments reveal that synchronization emerges when the state-dependent delay velocity coefficient reaches a threshold of \(\alpha \approx 5.308\). While Fig.~\ref{Fig.3}(c) (\(\alpha = 5\)) shows only a slight shift of \(x(t)\) towards the external forcing, Fig.~\ref{Fig.3}(e) (\(\alpha = 5.4\)) displays enhanced coherence, and Fig.~\ref{Fig.3}(g) (\(\alpha = 15\)) exhibits near-perfect overlap of the curves. This progression underscores the pivotal role of delay feedback strength in modulating synchronization within neural-like oscillatory systems.  Then, the corresponding phase portraits in the $xy$ plane are shown in Figs.~\ref{Fig.3}(b), (d), (f) and (h).  From the analysis of the figure, one can see that when $\alpha$ grows larger, the neuron's firing pattern becomes a periodic mixed-mode with an approximately constant amplitude. This means the state-dependent time-delay velocity component influences the firing pattern.

When $\alpha=-0.1$ the dynamics of the system are shown in Fig.~\ref{Fig.5}(a) and the response amplitude curve $Q$-factor in Fig.~\ref{Fig.5}(b) with a dot line. The curve shows a mixed/irregular firing patterns for $0 < g < 60$. Then, when the $g$ values grow the figure shows different transitions from less to more synchronized oscillations. The irregular firing patterns observed in the signal are associated with the potential effects of external electrical radiation on neuronal systems. This is the effect of using a sinusoidal external forces in our FitzHugh-Nagumo model. This choice is motivated by the ability of sinusoidal signals to represent periodic influences, such as those induced by electromagnetic fields or artificial electrical stimulation. These currents offer a controlled and analytically tractable method for studying the model's response to external stimuli. 

By introducing high- and low-frequency sinusoidal currents, we aim to capture resonance phenomena like vibrational and delay-induced resonance within a biologically plausible framework. While the primary focus of this study is theoretical, the results provide insights into how neural dynamics can be influenced by periodic environmental or artificial electrical forces, shedding light on the interplay between external perturbations and intrinsic neuronal behavior.

As the $\alpha$ parameter increases, one local maximum amplitude for $100 < g < 150$ becomes higher. In fact, for $\alpha=-0.01$ and $\alpha=0.13,$ the response amplitude $Q$ the local maximum grows.

Many slow-fast systems can exhibit state-dependent time-delay bifurcation, which means that the crucial transition occurs after some delay during the transition between the oscillatory and steady states because of the presence of a slowly varying parameter. Hence, in Fig.~\ref{Fig.6}, we specifically analyze the $Q$-factor and  the dynamical behavior of bifurcation state-dependent time delay FitzHugh–Nagumo neurons for values of amplitude $g$, in accord with Fig.~\ref{Fig.5}, before the vibrational resonance peak and on the peak. 
In our analysis, we include the exploration of the FitzHugh-Nagumo model with small negative values of the velocity delay coefficient \(\alpha\) to enhance the comprehensiveness of our study. Although positive \(\alpha\) represents constructive feedback mechanisms that amplify specific dynamics, negative \(\alpha\) introduces an opposing or suppressive feedback regime. This dual consideration enables the model to capture a wider spectrum of neuronal feedback behaviors, including inhibitory processes that are often observed in biological systems. In fact, negative \(\alpha\) can mimic mechanisms such as inhibitory synaptic feedback or adaptive processes that act to counterbalance excitation, thus maintaining stability or preventing excessive oscillatory behavior. Furthermore, from a dynamical systems perspective, the inclusion of negative \(\alpha\) probes the symmetry and stability of the system under varying feedback influences. This comprehensive analysis highlights how constructive and suppressive feedback mechanisms interact with state-dependent time delays to shape the observed vibrational and delay-induced resonance phenomena.

Furthermore, Fig.~\ref{Fig.6} is shown to illustrate the effect of the amplitude $\alpha$ of the state-dependent time-delay velocity component on the response amplitude $Q$ and on the dynamics of the system. In fact, we have plotted the response amplitude and the maxima and minima diagram as functions of the parameter $\alpha$ for different values of the amplitude $g$ corresponding to the high-frequency force. In Fig.~\ref{Fig.6}(a), when $g=10,$ there is a jump from being trapped in one well to oscillate around $x=0$ when $\alpha\approx 5$ and then it returns to its own well. On the other hand, in Figs.~\ref{Fig.6}(c) and (e) the oscillator jumps from oscillate around $x=0$ to one of the wells at $\alpha\approx 2$ and never goes back. Finally, in Figs.~\ref{Fig.6}(f) when $g=120$, the oscillation amplitude reach both wells and even if the variation of $\alpha$ has no effect on the qualitative dynamics of the system, it has on the oscillation amplitudes. In Figs.~\ref{Fig.6} (b), (d) and (f), the $Q$-factor plots do not show any sign of a resonance occurring, on the contrary they show an inverse peak localized at the $\alpha$ value for which the oscillator undergoes a change in its dynamics. Interestingly, in Figs.~\ref{Fig.6}(g) and (h), where $g=120$ coincides with the peak in Fig.~\ref{Fig.5}, the maxima and minima diagram and the $Q$-factor show a peak at $\alpha\approx 10$. This means that those specific values of the effect  of the $\alpha$ parameter on the synchronization of neuron firing pattern can give a boost to the oscillation amplitudes when the vibrational resonance is onset.

In Fig.~\ref{Fig.6b}, we show the gradient plot of the oscillation amplitudes and the response $Q$-factor as a function of the intensity $\alpha$ of the velocity state-dependent time delay component and the amplitude $g$ of the high-frequency force component. A study for different values of $p$ have been done and the results are qualitatively similar. These curves summarize all the values of the parameters $\alpha$ and $g$ for which the vibrational resonance phenomenon occurs. 

Here, we focus mainly on the case where \(\Omega = 10\omega\), but we also explored other combinations of frequencies, such as \(\Omega = 100\) with \(\omega = 1\) and \(\Omega = 10\) with \(\omega = 0.1\). These additional configurations align with the established principle that vibrational resonance arises when the high-frequency signal operates on a much faster time scale than the low-frequency signal.

Importantly, the nature of resonance does not universally result in a single resonance peak for all parameter configurations. The dynamics are influenced by the interaction between forcing amplitudes, system parameters, and the relative time scales of the signals. Although \(\Omega \gg \omega\) is a necessary condition, specific outcomes are contingent on these factors. We selected \(\Omega = 10\omega\) to explore a regime where the slowly-varying signal evolves on a sufficiently longer time scale than the rapidly oscillating signal, a critical feature for achieving vibrational resonance.

\section{The effect of the state-dependent time-delay velocity component on the delay-induced resonance}\label{Sec.4}

Delay-induced resonance presents a compelling extension to the concept of vibrational resonance by illustrating how intrinsic oscillations, generated by state-dependent time delays, can effectively replicate the role of high-frequency external forcing seen in classical vibrational resonance. These delay-induced oscillations, when coupled with a lower-frequency external periodic forcing, interact to produce resonance phenomena comparable to classical vibrational resonance mechanisms. Delay-induced resonance can thus be viewed as a special case of vibrational resonance, where oscillations arising due to the time delay in the system act as a high-frequency input interacting with external periodic forcing to generate resonance.

This perspective is supported by Lv et al.~\cite{lv2015high}, who demonstrated that delays can substitute for explicit high-frequency forcing in certain systems, making delay-induced oscillations an intrinsic contributor to resonance. By broadening the vibrational resonance framework, this approach shows that high-frequency periodic forcing is not indispensable; delays themselves can fulfill this role. In particular, delay-induced resonance occurs when oscillations of the system, induced by the delay, resonate with the forcing frequency, resulting in higher amplitude oscillations \cite{Cantisan,Coccolo}. Here, we analyze the parameter values for which this phenomenon arises in the FitzHugh-Nagumo model due to the state-dependent time delays position component, as well as the effects of the time delay velocity component on this resonance. To facilitate this analysis,  Eq.~(\ref{eq8}) is rewritten without the fast oscillating forcing term $g\cos\Omega t$, resulting in the following expression:
\begin{equation}
\begin{array}{lcl}
\ddot{x}+\mu
\left(1-0.54x+0.57 x^{2}\right)
\dot{x}+\displaystyle\frac{dV(x)}{dx}+ \gamma x\left(t-\tau(x(t))\right)+\alpha\dot{x}\left(t-\tau(x(t))\right)\\
\\
\quad \quad =f\cos\omega t.
\end{array}
\label{eq_delay_res}
\end{equation}
Note that this means $g=0$ in Eq.~(1).  Our study continues by analyzing the interaction between the state dependent time-delay position component and the remaining forcing. Thus, we can establish the parameters values for which the delay-induced resonance arise. Then, we will switch on the state dependent time-delay velocity component to observe its effect on the phenomenon.
Therefore, to start we set $\alpha=0$, so that the gain in amplitude is due just to the interaction between the state dependent time-delay position component $\gamma$ and the forcing $f\cos{(\omega t)}$. Hence, Fig.~\ref{Fig.7}(a) depicts the $Q-$factor in the parameter set $\omega-\gamma$ for $\tau_0=0.05$ and $p=-1$. It is possible to appreciate a yellow region and a tail that crosses the entire set that indicates where the delay-induced resonance takes place. In Fig.~\ref{Fig.7}(b), we show five slices along the $\omega$ axis of the previous figure for given $\gamma$ values and until $\omega=4$ for a better understanding.  The two first slices corresponding to $\gamma=0.64$ and $\gamma=1$, cross the yellow region. The third one is where $\gamma=1.44$ crosses the interaction of the tail and the yellow region. The last two for which  $\gamma=2$ and $\gamma=3$,  have been chosen to cross just the tail. In the five cases, the delay-induced resonance peak is observed, although displaced among each other. In fact, the curves exhibit varying shapes, shifting towards higher values of $\omega$ and decreasing in height as $\gamma$ increases. The curve associated with the smallest value of $\gamma$ shows three resonance peaks. The two higher peaks correspond to intrawell orbits, while the smaller peak corresponds to orbits confined within the right well. For all other values of $\gamma$, the resonance peaks indicate intrawell orbits. 

Additionally, there is a noticeable transition from a more abrupt emergence of the peak to a smoother variation as the value of $\gamma$ increases. In Fig.~\ref{Fig.7}(b), the values $\gamma=1.44$ and $\gamma=2$ mark two significant changes in the behavior of the curves. The first change is that the resonance peak diminishes. The reason for that is due to the fact that the $\gamma$ value is intercepting the last part of the yellow region and the beginning of the tail. The second change is the transition from a sudden shift in the curves' behavior to a more gradual variation of the $Q$ curve. Although the curves in Fig.~\ref{Fig.5} and Fig.~\ref{Fig.7}(b) share a certain resemblance and the peak means the birth of a resonance they are associated to different phenomena. In fact, in the first case, the resonance is triggered by the interaction between the two external forcing terms. In the second case, one of the forcing term has been switched off. In this way, we can be sure that the resonance comes from the interplay between the time-delay position component and the remaining external forcing. The behaviors of the system are depicted in Fig.~\ref{Fig.7}(c-f). 

\begin{figure}[htp]
\begin{center}
\includegraphics[width=16cm, ,clip=true]{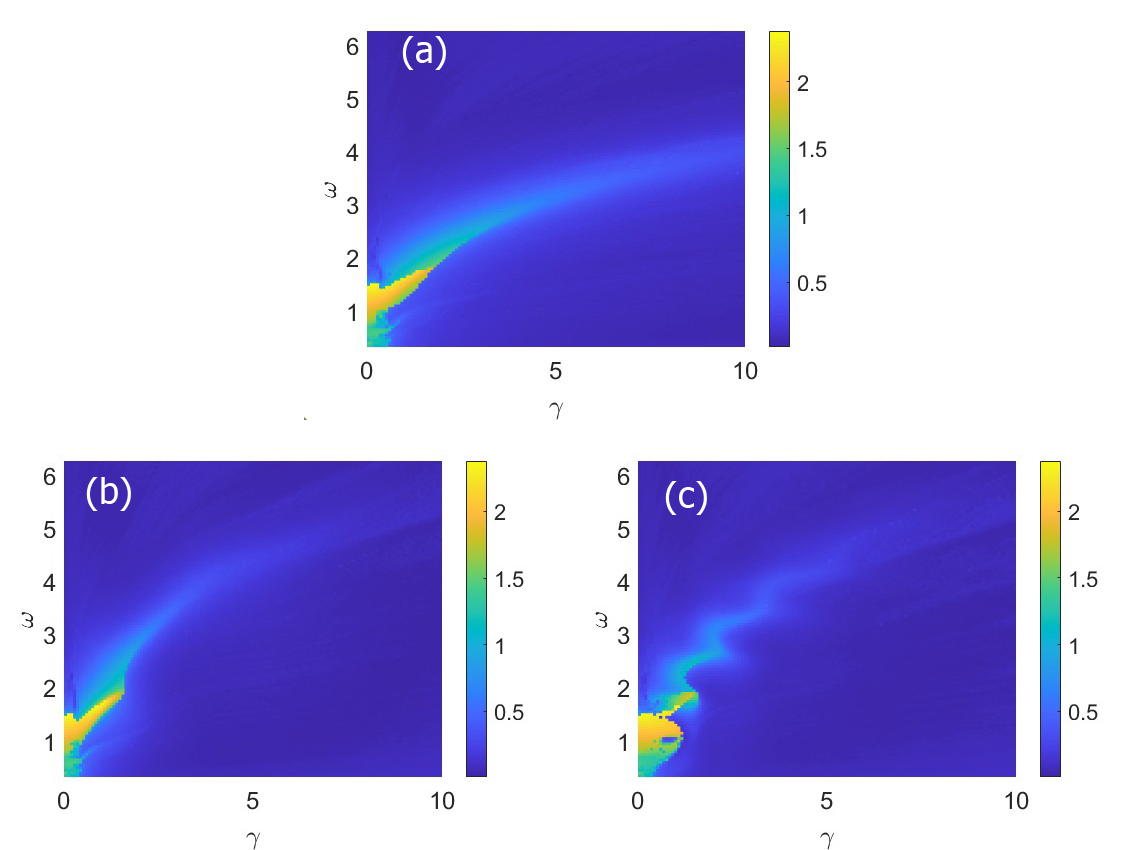}
\caption{The gradient of the $Q-$factor indicates the presence of delay-induced resonance in the FitzHugh-Nagumo model with a state-dependent sigmoidal function time delay. For this analysis, we have fixed the values in the parameter set of $\gamma$ the multiplier of the velocity delay and $\omega$ the frequency of the forcing $f\cos{(\omega t)}$. The other parameters are: $\mu=0.1$, $g=0$, $\alpha=0$, $f=1$,  $p=-1$, and $\tau_{0}=1.5$ in panel (a), $\tau_{0}=3.5$ in panel (b),  $\tau_0=7.5$ in panel (c). We observe that for values of $\tau_0<3.5$, the results are qualitatively the same, with minimal changes in the area of the yellow region compared to the case presented in Fig.~\ref{Fig.7}. This indicates that the phenomenon is robust against variations in the $\tau0$ parameter.} \label{Fig.8}
\end{center}
\end{figure}

Next,  we analyze the effect of the sigmoid function parameter $\tau_0$  on the phenomenon. We observe that for  $\tau_0<3.5$, its impact is negligible, as the differences between Fig.~\ref{Fig.8}(a) and Fig.~\ref{Fig.7}(a) are minimal. However, for larger values of $\tau_0$, the effects become more pronounced, as illustrated in Figs.~\ref{Fig.8}(b) and~\ref{Fig.8}(c). In these last figures, the tail changes shape at $\tau=3.5$ and begins to spin multiple times in the set space for larger $\tau$ values. We have also examined the impact of the $p$ parameter on delay-induced resonance. This parameter does not significantly affect the phenomenon until we approach the threshold value of $\tau_0 \approx 3.5$ and even beyond this $\tau_0$ threshold, the influence of $p$ remains minimal. Therefore, without loss of generality, we will fix $p=-1$ for the rest of our analysis.

\begin{figure}[htp]
\begin{center}
\includegraphics[width=16cm, ,clip=true]{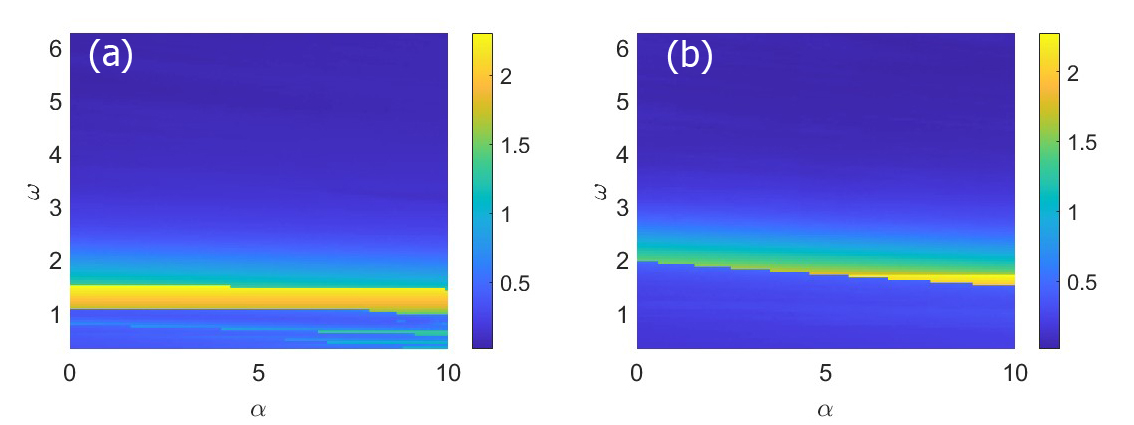}
\caption{The gradient of the $Q-$factor identifies the presence of delay-induced resonance in the FitzHugh-Nagumo model with a state-dependent sigmoidal function time delay. We have fixed the values in the parameter set of $\alpha$ the multiplier of the time-delay velocity component and $\omega$ the frequency of the forcing $f\cos{(\omega t)}$. The other parameters are: $\mu=0.1$, $g=0$, $f=1$,  $p=-1$, $\tau_{0}=0.05,$.  The multiplier of the position delay is $\gamma=0.8$ (a) and $\gamma=2$ (b). In the first case (a), with $\gamma=0.8$, delay-induced resonance is maintained for all $\alpha$ values, resulting in only minor changes to the yellow area. In contrast, in the second case (b), where $\gamma=2$ is outside the yellow region of Fig.~\ref{Fig.7} (a), delay-induced resonance can be triggered over a broader spectrum of $\omega$ values when $\alpha$ becomes sufficiently strong. } 
\label{Fig.9}
\end{center}
\end{figure}

After identifying the parameter values that trigger delay-induced resonance, we introduce a non-zero time-delay velocity component ($\alpha > 0$) to examine its effect on the phenomenon. From this analysis we obtain Fig.~\ref{Fig.9}, that displays the $Q$-factor in the $\omega-\alpha$ parameter set for two different values of $\gamma$ and using the same parameter values as in Fig.~\ref{Fig.7}(a) except for $\alpha$. Specifically, in Fig.~\ref{Fig.9}(a), we set $\gamma=0.8$, which is within the yellow region indicating a delay-induced resonance in Fig.~\ref{Fig.7}. In Fig.~\ref{Fig.9}(b), $\gamma$ is set to $2$, a value outside this region but intersecting the tail of the delay-induced resonance at $\omega \approx 2$ in Fig.~\ref{Fig.7}(a). The first panel shows that changing the velocity delay component does not significantly impact the phenomenon. However, the second panel demonstrates that the time-delay velocity component can expand the range of $\omega$ values that trigger resonance for larger $\alpha$ values. For smaller $\alpha$ values, resonance is triggered at $\omega \approx 2$, as seen in Fig.~\ref{Fig.7}(a). Conversely, when $\alpha \gtrsim 5$, the yellow region expands along the $\omega$ axis, indicating the onset of a resonance.  This means that the addition of the state dependent time-delay velocity component, enhancing the synchronization between the neuron and external driving signals, can control the appearance of the resonance phenomenon. 

The results of this section provide valuable insights into the dynamic interplay between time delay and periodic forcing in the FitzHugh-Nagumo neuron model. By introducing state-dependent time delays into both the position and velocity terms, we uncover a nuanced mechanism through which resonance phenomena are influenced. This exploration bridges traditional resonance and delay-induced effects, demonstrating how delays can substitute or complement high-frequency external forcing in generating vibrational resonance.

Furthermore, the outcomes highlight the sensitivity of the system to variations in the delay parameters and forcing frequencies, revealing a rich structure of resonance regions. These findings emphasize the importance of considering delays as intrinsic components of the system, not just external influences, and expand the conceptual framework of vibrational resonance to encompass state-dependent effects.

By establishing that delay-induced oscillations can fulfill roles akin to high-frequency periodic forcing, we propose a broader interpretation of vibrational resonance phenomena. This perspective is not only theoretically significant but also potentially impactful in understanding biological or physical systems where delays naturally arise, such as neural networks or coupled oscillatory systems.

\section{Conclusions}\label{Sec.5}

Initially, we have presented and discussed the effect of the state-dependent time-delay velocity component on vibrational resonance. Our results highlight the importance of incorporating an asymmetric double-well potential in our physical model, as it leads to various outcomes. Specifically, under the combined influence of the high-frequency amplitude and the intensity of the state-dependent time-delay component, we have demonstrated the possibility of vibrational resonance in the neuronal model. However, we have observed that an increased intensity of the state-dependent time-delay component tends to reduce the amplitude of the vibrational resonance. Additionally, vibrational resonance may not manifest for high ranges of the parameter $\alpha.$

Later, we have examined the emergence of delay-induced resonance in the FitzHugh-Nagumo system and the influence of the delay velocity component on this phenomenon. We found that a specific set of values for the delay position and the frequency of the external forcing, $f\cos(\omega t)$, is crucial for initiating this phenomenon. Furthermore, we identified two scenarios regarding the effect of the delay velocity component. First, if the delay-induced resonance has already been triggered, the delay velocity does not influence it. Second, when the delay position component activates the phenomenon for a specific forcing frequency, the delay velocity can broaden the spectrum of frequencies that can induce it.

In summary, we can conclude that the state-dependent time-delay velocity component plays a crucial role in controlling the appearance of vibrational resonance, and delay-induced resonance.

\section{Acknowledgment}

This work was supported by the Spanish State Research Agency
(AEI) and the European Regional Development Fund (ERDF, EU)
under Project Nos. PID2019-105554GB-I00 and PID2023-148160NB-I00 (MCIN/AEI/10.13039/501100011033). It has also been supported by Rey Juan Carlos University, Spain, through the Programa Propio de Fomento y Desarrollo de la Investigación de la URJC under the project NABEH, funded by Financiación para proyectos Impulso (Project No. 2024/SOLCON-137648).

\newpage

 \end{document}